\documentclass[%
 reprint,
 amsmath,amssymb,
 aps,
]{revtex4-1}
\usepackage{xcolor}
\usepackage{graphicx}
\usepackage{dcolumn}
\usepackage{bm}

\DeclareMathOperator\erfc{erfc}

\begin{document}


\title{Density Functional Theory Formulation for Fluid Adsorption on \\
	 Correlated Random Surfaces
	}

\author{Timur Aslyamov}%
 \email{t.aslyamov@gmail.com; taslyamov@slb.com}
 \affiliation{%
 	Schlumberger Moscow Research Center; 13, Pudovkina str., Moscow 119285, Russia}

\author{Aleksey Khlyupin}
 \email{khlyupin@phystech.edu}
\affiliation{%
Moscow Institute of Physics and Technology; Institutsky lane 9, Dolgoprudny, Moscow region, 141700, Russia
\\
Schlumberger Moscow Research Center; 13, Pudovkina str., Moscow 119285, Russia 
}%
\date{\today}
\begin{abstract}
We provide novel random surface density functional theory (RSDFT) formulation in the case of geometric heterogeneous surface of solid media which is essential for description of thermodynamic properties of confined fluids. The major difference of our theoretical approach from existing ones is stochastic model of solid surface which takes into account the correlation properties of geometry. The main building blocks are effective fluid-solid potential developed in work (J. Stat. Phys, 2017,167(6), 1519-1545) and geometry based modification of Helmholtz free energy for Lennard-Jones fluids. Efficiency of RSDFT is demonstrated in calculation of argon and nitrogen low temperature adsorption on real heterogeneous surfaces (BP280 carbon black). These results are in good agreement with experimental data published in the literature. Also several models of corrugated materials are developed in the framework of RSDFT.  Numerical analysis demonstrates strong influence of surface roughness characteristics on adsorption isotherms. Thus developed formalism provides connection between rigorous description of stochastic surface and confined fluids thermodynamics.
\end{abstract}

\pacs{Valid PACS appear here}
\maketitle


\section{Introduction}
Real surfaces are usually rough, so influence of geometry on adsorption and other surface properties is actively investigated in recent years \cite{Khlyupin2017, ravikovitch2006density, neimark2009quenched, jagiello2013carbon, jagiello2015dual, do2006modeling, do2005gcmc,
	quere2002rough, netz1997roughness, coasne2013adsorption, herminghaus2012universal, yatsyshin2017classical, persson2004nature, forte2014effective, ustinov2006pore, ustinov2005application, khlyupin2016effects}. A huge range of surface phenomena can be described using Density Functional Theory (DFT) \cite{wu2006density}. Especially DFT plays essential role in theoretical prediction of adsorption characteristics. However the major part of known versions of DFT is applicable for adsorption on smooth surfaces only. 

In order to avoid assumptions about smooth solid surface in the frame of one-dimensional DFT authors of 
\cite{ ravikovitch2006density, neimark2009quenched} considered the heterogeneous  material as an amorphous media with variable one-dimensional density near the surface. Therefore the solid density is represented as rapidly decreasing function of the distance to the surface. Thus, the heterogeneity is described by a single parameter corresponding to the value of characteristic roughness. In accordance with this representation of solid in work \cite{ravikovitch2006density} new version of DFT named the quenched solid density functional theory (QSDFT) was developed. Results of QSDFT fit experimental data in the range of thermodynamic parameters where results of standard DFT are insufficient. However, QSDFT has serious limitation due to simplified representation of solid surface. Indeed single roughness parameter is not enough for full definition of heterogeneous surface and the characteristic of lateral structure is needed. Alternatively authors of \cite{jagiello2013carbon, jagiello2015dual} considered two-dimensional heterogeneous surface, taking into account that carbon structure consists of curved graphene layers. In order to describe lateral surface energy perturbations authors introduce deterministic oscillating function.  However more realistic model of rough solid is random surface defined by both characteristic roughness and correlation function. Thus new DFT formulation supporting the analysis of influence caused by correlation properties of the random solid surface is highly desirable. In the current manuscript we present the theory which successively describes the heterogeneous surface and real fluid thermodynamics near the wall in common formalism.

In the previous work \cite{Khlyupin2017}, we developed the general theory of effective coarse-grained fluid-solid potential by proper averaging of the free energy of fluid molecules which interact with the random solid media. This procedure is largely based on the theory of random processes so we focused on a detailed calculation of the averaged local geometry properties using the random field approach. Correlated random field is considered as a model of random surface with high geometric roughness. As a result, general expression of effective fluid-solid potential was obtained and discussed in details.

\begin{figure*}
	\centering
	\includegraphics[width=16.5cm]{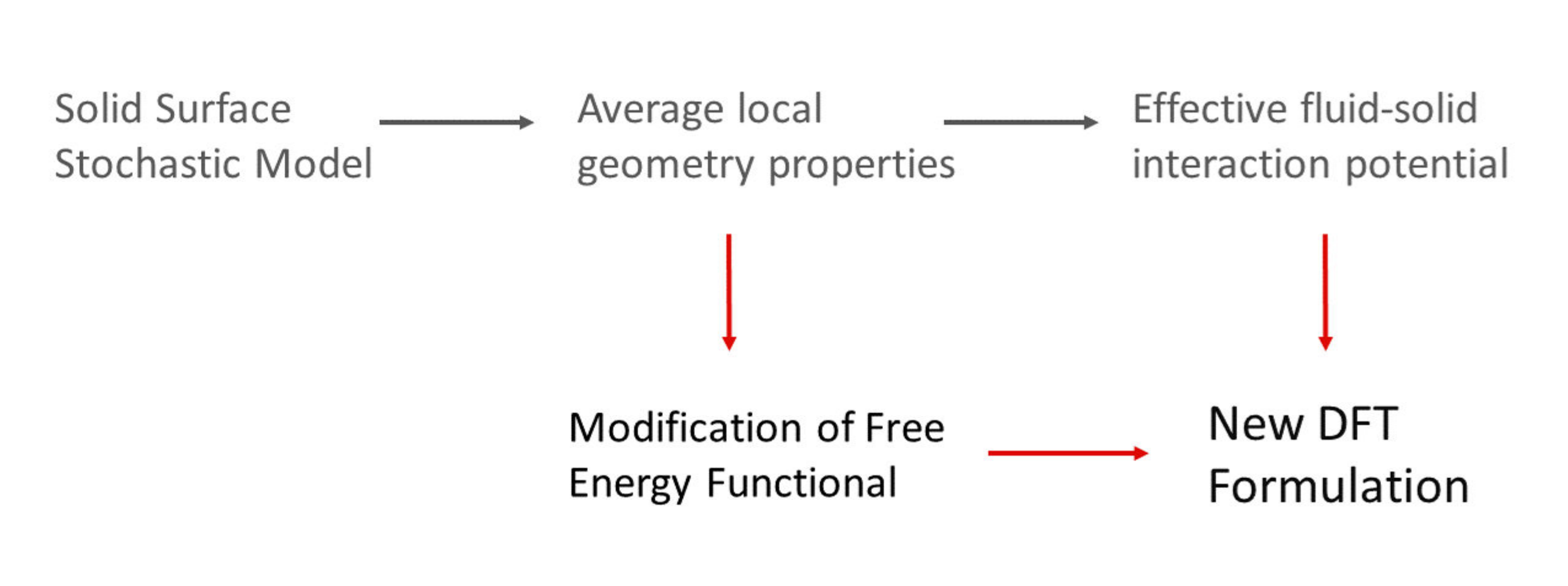}
	\caption{\label{SchemeWorkFlow} Main building blocks of the new DFT formulation scheme. Gray lines demonstrate stages which were developed and discussed in details in our previous work \cite{Khlyupin2017}. Red lines correspond to current research results.}
\end{figure*}

In the current paper we have developed a new density functional theory to describe thermodynamic properties of a fluid interaction with strongly spatial heterogeneous surfaces. One of the building blocks for creating such a theory is description of the effective fluid-solid interaction taking into account the properties of the stochastic surface geometry Fig.~\ref{SchemeWorkFlow}.  Also the averaged properties of the local solid geometry underlying the theoretical model are essential for the modification of the configuration integral in free energy functional. This procedure  is described in details in the current manuscript. Collecting all this together, we construct a new density functional theory on the basis of unified approach, which implies accurate consideration of both single-point and two-point distribution functions of a random surface.

\begin{figure*}[tp]
	\centering
	\includegraphics[width=16.5cm]{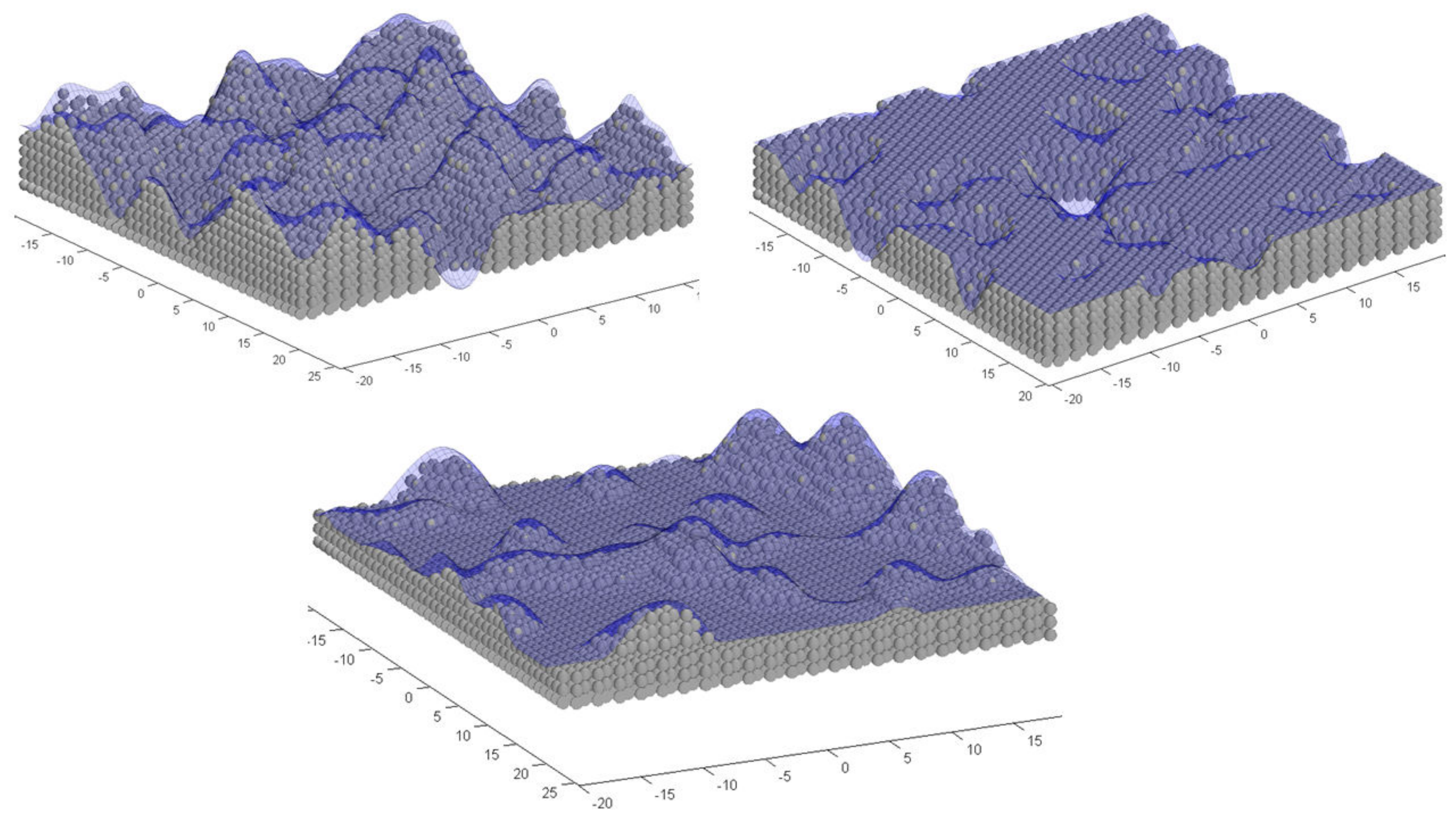}
	\caption{\label{IntroCorrugated}
		Schematic illustration of three different models of corrugated heterogeneous materials. Gray balls illustrate solid molecules. Solid surface can be described by certain realization of the random fields: Gaussian correlated random field (left), truncated Gaussian random field from above (right), truncated Gaussian random field from below (middle). See text for details.}
\end{figure*}

Also, in this study, we have demonstrated how the developed formalism may be expanded to describe non-symmetric models of random solid surface. This allows us to consider two types of corrugated materials. The first one can be obtained by chemical reactions in which random  cavities are formed in the initially smooth surface. The second one is devoted to random geometry formed by depositing solid matter onto originally smooth substrate, illustration of different surface types which have been studied is provided in  Fig.~\ref{IntroCorrugated}. 

In work \cite{do2006modeling} nongraphitized carbon black (NGCB) surfaces were investigated by modeling adsorption of argon on these surfaces in the framework of fully atomistic Grand Canonical Monte-Carlo simulation (GCMC). The new simple model to describe a surface irregularities of corrugated materials was presented since many real surfaces are far from being ideally smooth, and assuming a perfect surface to study adsorption in pores could lead to serious errors in the determination of adsorption isotherms.  Also in paper \cite{do2006modeling} authors concentrate on comparison of adsorption behaviors between NGCB and graphitized thermal carbon black (GTCB) with well defined smooth surface. The models of heterogeneous surfaces constructed in our manuscript have many similarities with the model from work \cite{do2006modeling} and expand it within the framework of the developed theoretical approach. This paper is arranged as follows: necessary definitions and results of our previous research are briefly discussed in the next section. Then we formulate new version of DFT named Random Surface Density Functional Theory (RSDFT). The results are presented in two sections. The first one demonstrates RSDFT application to low temperature  adsorption on nongraphitized carbon black BP-280. In this section results obtained theoretically are compared with experimental data. The second one demonstrates new approach to modeling of various types of corrugated materials.

\section{Random Process Theory Approach to Geometric Heterogeneous Surfaces}

Since the model developed in our previous work \cite{Khlyupin2017} is much in use in the current manuscript, let us briefly remind main results and constructions. The underlying idea states that total interaction between fluid molecule and solid media can be replaced by an effective potential depending on the distance from a molecule to the surface only. Implemented free energy average technique  is an approach to link the free energy, obtained by accurate partition function of initial system, to a system with fewer degrees of freedom which is result of approximated partition function.  

Let us consider a molecule of fluid, interacting with solid phase. Fluid molecule is a sphere with diameter $d$. The solid is represented by a system of $M$ noninteracting molecules located at the sites of three dimensional lattice. Solid media has sufficiently large surface, which corresponds to certain realization of random process $\mathcal{Z}(r)$. For a fluid particle with fixed z-coordinate corresponding partition function becomes

\begin{eqnarray}
	\label{PartFunction(2)}
	Q(z;\mathcal{Z}(r))=\iint\limits_{\Omega(z)}dx_fdy_f e^{-\beta \sum_{i=1}^{M}u_{fs}\left(\vec{r}_f,\vec{r}_s^{(i)}\right)}
\end{eqnarray}

The configuration space of fluid molecule in expression \eqref{PartFunction(2)} is union of non-crossing domains of random sizes,  $\Omega(z)=\bigcup_{i=1}^{\infty}\Omega_i(z)$. Permitted random regions $\Omega_i$ are induced by random binary field which is a slice of random process $\mathcal{Z}$ at level $z$, $u_{fs}(\vec{r}_f,\vec{r}_s^{(i)})$ is pair interaction potential between fluid molecule at point $r_f$ and solid molecules at fixed $M$ points $r_{s,1},r_{s,2},...,r_{s,M}$. It is important to note, that this function also depends on random process $\mathcal{Z}(r)$.  Thermodynamic properties of this system are defined by free energy averaged over all realizations of random geometry:

\begin{eqnarray}
	\label{Aver_F_def(5)}
	\beta \left\langle F(z)\right\rangle_\mathcal{Z}\equiv-\int\ln Q(z;\mathcal{Z}(r))P(\mathcal{Z})\prod_r d\mathcal{Z}(r) 
\end{eqnarray}
where integrals imply functional integration over variations of $\mathcal{Z}$ at each point $r$. $P(\mathcal{Z})$ is probability that certain $\mathcal{Z}$ takes place. As it was shown \cite{Khlyupin2017}, effective potential of interaction between fluid molecule and solid with random heterogeneous surface has the following form:
\begin{eqnarray}
	\label{EffPotential_2(12)}
	&U_{fs}^{eff}(z)=\int dr_s u_{fs}(r_f, r_s) \times \\ \nonumber
	&\times \int d\rho(r_s)\rho(r_s)P\left(\rho(r_s)|r_f\in\bar{\Omega}(z)\right)
\end{eqnarray}
where $P\left(\rho(r_s)|r_f\in\bar{\Omega}(z)\right)$ is probability density of $\rho(r_s)$ under condition that fluid particle lies in characteristic domain $\bar{\Omega}(z)$. Thus, effective fluid-solid potential reflects the random surface properties by probability density $P$ and average size of $\bar{\Omega}$. Here the solid density is introduced as the number of solid molecules in volume $d\vec{r}_s$.  $\bar{\Omega}(z)$ is domain with the average (characteristic) size. This average size depends on the properties of random process and the coordinate $z$. 

\begin{figure}[tp]
	\includegraphics[width=8.5cm]{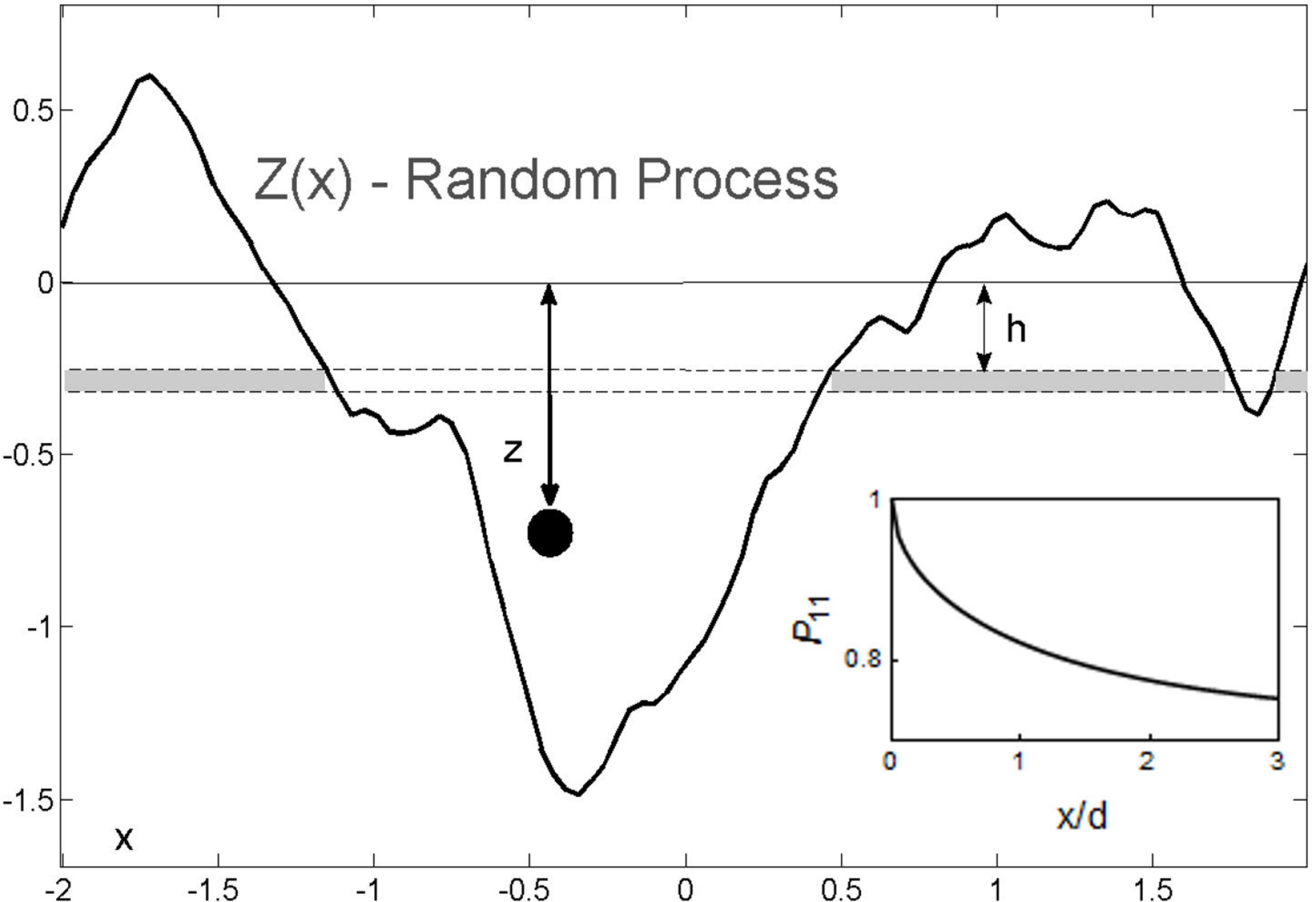}
	\caption{\label{ProcessExample} 
		 Illustration of the molecule (black disk) and some realization of random process corresponding to heterogeneous surface (solid curve). In this case the slice at level $h$ is above the fluid particle. Inset shows typical form of $P_{11}$, calculation parameters are $\alpha=0.6 d^{-1}, \sigma=0.5 d, h=-0.3 d$}
\end{figure}

Gaussian process is considered as basic model of random surface. Thus, at this case one-point $w_z^{(1)} (x)$  and two-points density distribution function $w_z^{(2)}(z_1,x_1;z_2,x_2 )$  with arbitrary correlation function $K(s)$ have the following forms

\begin{eqnarray}
	\label{Gauss_dens(21)}
	w_z^{(1)}(z)=\dfrac{1}{\sqrt{2\pi\sigma^2}}\exp\left(-\dfrac{z^2}{2\sigma^2}\right)
\end{eqnarray}
\begin{eqnarray}
	\label{Gauss_distr(19)}
	w_z^{(2)}(z_1, z ;z_2, z+s)=\dfrac{1}{2\pi\sigma^2 \sqrt{1-K(s)^2}}\times \nonumber \\
	\times\exp\left[-\dfrac{z_1^2+z_2^2-2K(s) z_1z_2}{2\sigma^2(1-K(s)^2)}\right]
\end{eqnarray}

The theoretical approach may be applied for arbitrary correlation function of the random solid surface which could be obtained from experiments (for example X-ray measurements). However exponential function of the form $K(x)=e^{-\alpha x}$, (where $\alpha=1/\tau$  is inverse to correlation length $\tau$),  is convenient for the calculations and demonstrates the general properties.  Important to note, that the approach proposed in \cite{Khlyupin2017} is not restricted by certain functional dependence for distribution functions. Gaussian model was chosen only for simplicity. 

It is possible to obtain the general result for the effective fluid-solid interaction potential starting from  popular Mie pair intermolecular potential in the following form:
\begin{eqnarray}
	\label{Mie(51)}
	U(r)= C \left[ \left(\frac{d}{r}\right)^{\lambda_r}-\left(\frac{d}{r}\right)^{\lambda_a}\right],
\end{eqnarray}
where $r$ is the distance between molecules, $C=\frac{ \epsilon_{sf}\lambda_r}{\lambda_r-\lambda_a}\left(\frac{\lambda_r}{\lambda_a}\right)^{\frac{\lambda_r}{\lambda_r-\lambda_a}}$ is a constant. In case of Lennard-Jones (LJ) fluid ($\lambda_r=12, \lambda_a=6$) this constant equals to $C=4\epsilon_{sf}$, where $\epsilon_{sf}$ is characteristic energy of solid-fluid interaction. One can consider each term of above expression as general power function of $r$ in the following form $C d^{\gamma}/r^{\gamma}$. The interaction energy of molecule and surface, induced by pair potential $U(r)$ is the sum of interactions with all molecules in the solid media. Thus, integration of $U(r)$ over certain layer at level $h$ (for example, gray area in Fig.\ref{ProcessExample}) in z-direction becomes:

\begin{equation}
	\label{Heter_Pot_Layer_2(54)}
	U_{layer}^{(\gamma)}(\mathcal{L};h,z)=2\pi C\int_{0}^{\infty}\frac{dr'r'\rho_s P_{11}(r',h)}{\left[\left(z-h\right)^2+\left(r'+\mathcal{L}\right)^2\right]^{\gamma/2}}
\end{equation}
where $\rho_s$ is the number density of solid molecules. Expression \eqref{Heter_Pot_Layer_2(54)} is written in terms of average solid density $\rho_s P_{11}(r, h)$ and average length $\mathcal{L}(z)$ corresponding to averaged distance from molecules to solid surface at level $z$. Typical example of this function can be found in inset of Fig.~\ref{ProcessExample}. Exact expressions for average length $\mathcal{L}$ and $P_{11}$ were obtained in \cite{Khlyupin2017} using random process theory approach. In accordance with \eqref{Mie(51)} the impact of each layer at level $h$ has the following form
\begin{eqnarray}
\label{U_layer(55)}
&U_{layer}(\mathcal{L};h,z)=U_{layer}^{(\lambda_r)}(\mathcal{L};h,z)-U_{layer}^{(\lambda_a)}(\mathcal{L};h,z)
\end{eqnarray}

Total effective interaction of a molecule and the solid media corresponds to integration over all layers from $-\infty$ to $\infty$
\begin{eqnarray}
\label{external_field}
U_{fs}(z)=\int_{-\infty}^{\infty}dh U_{layer}(\mathcal{L};h,z)
\end{eqnarray}

\begin{figure}
	\includegraphics[width=8.5cm]{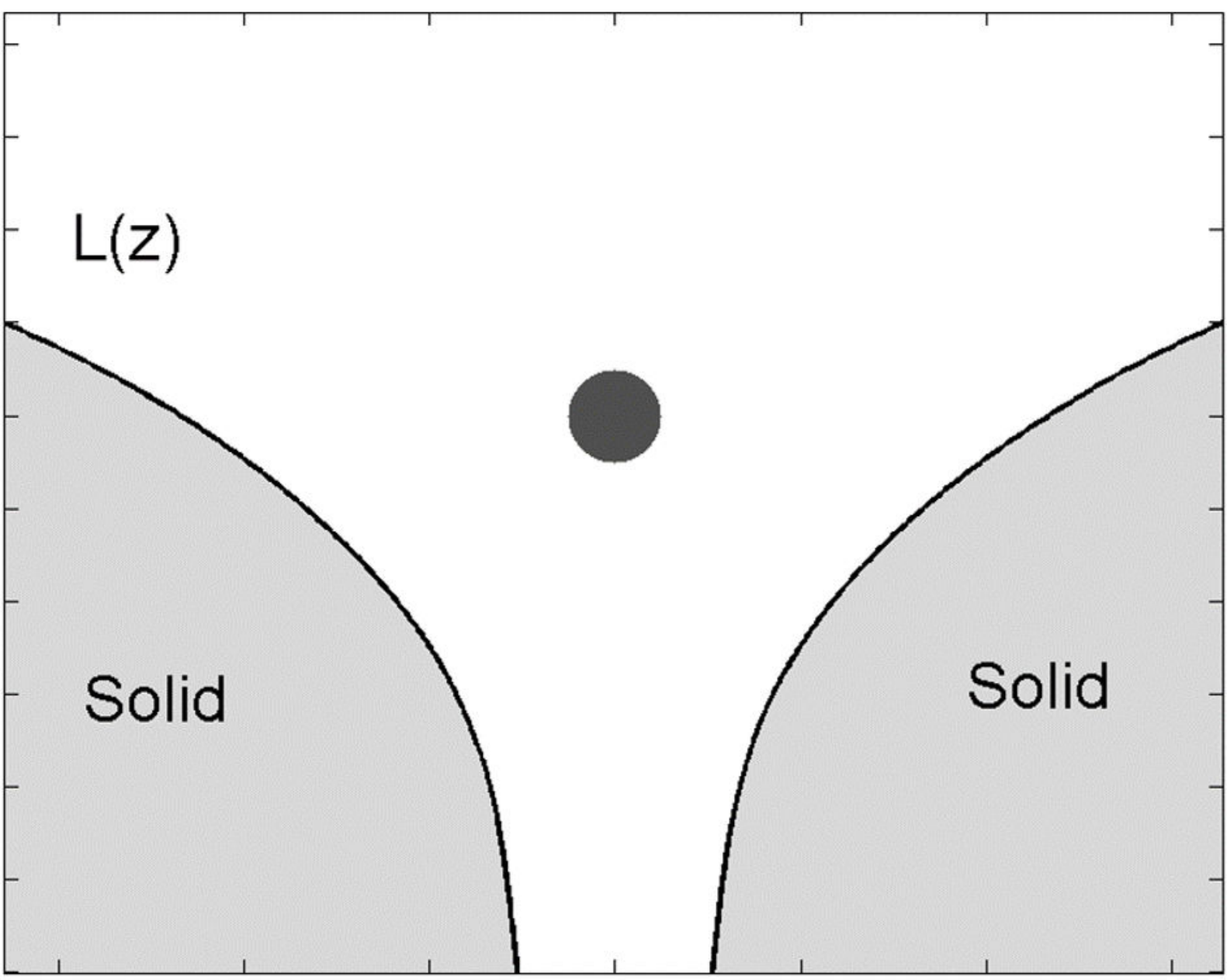}
	\caption{
		\label{Particle_at_z} Schematic illustration demonstrates the local surface geometry properties, which are described by function $\mathcal{L}(z)$.
	}
\end{figure}

The characteristic properties of the local geometry are reflected in an explicit form of the function $\mathcal{L}$ which is used to calculate the effective potential \eqref{U_layer(55)}. The knowledge of local geometry is essential not only for understanding the effective potential but also for modifying the configuration integral according to heterogeneous geometry of the surface. These modifications are required in the new formulation of the free energy functional.  Schematic illustration of molecule and solid surface in terms of function $\mathcal{L}$ can be found in Fig.~\ref{Particle_at_z}.  This figure illustrates that heterogeneous surface can be locally represented as the known surface defined by function $\mathcal{L}$. Thus new surface has the polar symmetry and defines boundary of excluded volume taking into account geometrical properties of the solid. 

\section{Random Surface Density Functional Theory}

This section discuses new DFT formulation -- Random Surface Density Functional Theory (RSDFT). As in the case of homogeneous surface the starting point of RSDFT is grand canonical thermodynamic potential $\Omega[\rho(\vec{r})]$:

\begin{eqnarray}
	\label{GCE}
	\Omega[\rho(\vec{r})]=F[\rho(\vec{r})]+\int_V d\vec{r}\rho(\vec{r})(U_{fs}(\vec{r})-\mu)
\end{eqnarray}  
where $F[\rho]$ is the Helmholtz free energy, $U_{fs}$ is the external potential, $\mu$ is the chemical potential. The equilibrium density distribution $\rho$ is satisfied to the following condition:
\begin{eqnarray}
	\label{Eq_1}
	\dfrac{\delta \Omega[\rho]}{\delta \rho}=0
\end{eqnarray} 

The Helmholtz free energy can be splitted into two parts: the ideal term $F_{id}$ and the excess term $F_{exc}$ describing contributions of intermolecular repulsion and attraction. Henceforth Boltzmann and Plank constants are assumed to be equal to one: $k_B=h=1$. Thus, Helmholtz free energy can be expressed as
\begin{eqnarray}
	F[\rho]=F_{id}[\rho]+F_{exc}[\rho]
\end{eqnarray}

\subsection{Ideal contribution}

For an ideal system without any interactions, the Helmholtz free energy is known exactly :
\begin{eqnarray}
	\label{F_id}
	F_{id}[\rho]=T \int_V d\vec{r}\rho(\vec{r})[\ln(\Lambda^3 \rho(\vec{r}))-1]
\end{eqnarray} 
where $T$ is the temperature, $\Lambda=(2\pi m T)^{1/2}$ is de Broglie wavelength, $m$ is the mass of molecule. Expression \eqref{F_id} contains three-dimensional integration which can be reduced to one-dimensional integral using spatial symmetry. As was discussed in previous section interaction with rough surface can be represented as impact of the solid media with polar symmetry Fig.~\ref{Particle_at_z}B. It allows one to introduce one-dimensional density $\rho(z)$. Contrary to case with smooth surface, result of integration $\int dxdy...$ is not constant. One can calculate this two dimensional integral taking into account only permitted domains for fluid  on certain level $z$ as 
\begin{eqnarray}
	\int dz\rho(z)\int_A dxdy...=\int dz\rho(z)S(z)...
\end{eqnarray}
where $A$ is total area, $S(z)$ is the part of $A$ which is free from solid media at level $z$. One can found this area from the following expression
\begin{eqnarray}
	\label{S(z)}
	S(z)=A\left(1-\dfrac{1}{2}\erfc\dfrac{z}{\sqrt{2}\sigma}\right)
\end{eqnarray}
from equation \eqref{S(z)} one can see that the cases $\sigma\to0$ and $z\to\infty$ are equivalent to the case of smooth surface $S(z)\to A$. Thus ideal part \eqref{F_id} has the following form:
\begin{eqnarray}
	\label{F_id_mod}
	F_{id}[\rho]=T \int dz S(z)\rho(z)[\ln(\Lambda^3 \rho(z)-1]
\end{eqnarray} 

\subsection{Attraction contribution}

Molecules are considered as not ideal system which  is represented by spherical molecules with diameter $d$ interacting via Lennard-Jones (LJ) potential $U_{LJ}$
\begin{eqnarray}
	U_{LJ}=4\epsilon_{ff}\left[\left(\dfrac{d}{r}\right)^{12}-\left(\dfrac{d}{r}\right)^6\right]
\end{eqnarray}
where $\epsilon_{ff}$ is the characteristic intermolecular energy.

The excess term can be calculated using perturbation theory for the interaction potential \cite{kalikmanov2013statistical}. The repulsion contribution can be described by reference system -- the system of hard sphere, while attraction contribution corresponds to the perturbation term. Thus, the excess Helmholtz free energy can be written as
\begin{eqnarray}
	F_{exc}[\rho]=F_{HS}[\rho]+F_{att}[\rho]
\end{eqnarray} 
Here for perturbed attraction part we use Weeks-Chandler-Andersen (WCA) scheme \cite{kalikmanov2013statistical} with the following representation of attraction potential
\begin{eqnarray}
	\label{WCA}
	U_{att}(r)=\begin{cases}
		-\epsilon, \,\,\, r<\lambda \nonumber \\
		U_{LJ}(r), \,\,\, r>\lambda 
	\end{cases}
\end{eqnarray} 
where $\lambda=2^{1/6}d$ corresponds to the minimum of LJ potential. The mean-field approximation for attraction part is
\begin{eqnarray}
	\label{F_att}
	F_{att}=\dfrac{1}{2}\int\int d\vec{r'}d\vec{r}\rho(\vec{r})\rho(\vec{r'})U_{att}\left(|\vec{r}-\vec{r'}|\right)
\end{eqnarray} 

The number of integrals in expression \eqref{F_att} can be reduced by the same procedure as for ideal term taking into account spatial dependence in \eqref{WCA}
\begin{eqnarray}
	\label{F_att_mod}
	F_{att}=\dfrac{1}{2}\int\int dz_1dz_2S(z_1)\rho(z_1)\rho(z_2)G(z_1,z_2)
\end{eqnarray}
where function $G(z_1,z_2)$ is defined as the following integral:
\begin{eqnarray}
	G(z_1,z_2)=2\pi \int_{0}^{\infty}r dr U_{att}\left(\sqrt{(z_1-z_2)^2+r^2}\right)
\end{eqnarray}

One can calculate this integral using the expression for effective media solid boundary $\mathcal{L}(z)$ at level $z$ and piecewise definition of $U_{att}$:
\begin{eqnarray}
	\label{Kernel}
	G(z_1,z_2)=\begin{cases}
		G_1(z_1,z_2), \,\,\, (z_2-z_1)^2>\lambda^2 \\
		G_2(z_1,z_2), \,\,\,\mathcal{L}(z_2)^2+(z_2-z_1)^2>\lambda^2  \\
		G_3(z_1,z_2), \,\,\, \mathcal{L}(z_2)^2+(z_2-z_1)^2 \leq \lambda^2 
	\end{cases}
\end{eqnarray}
explicit form of \eqref{Kernel} can be found in Appendix.

\begin{figure}
	\includegraphics[width=8.5cm]{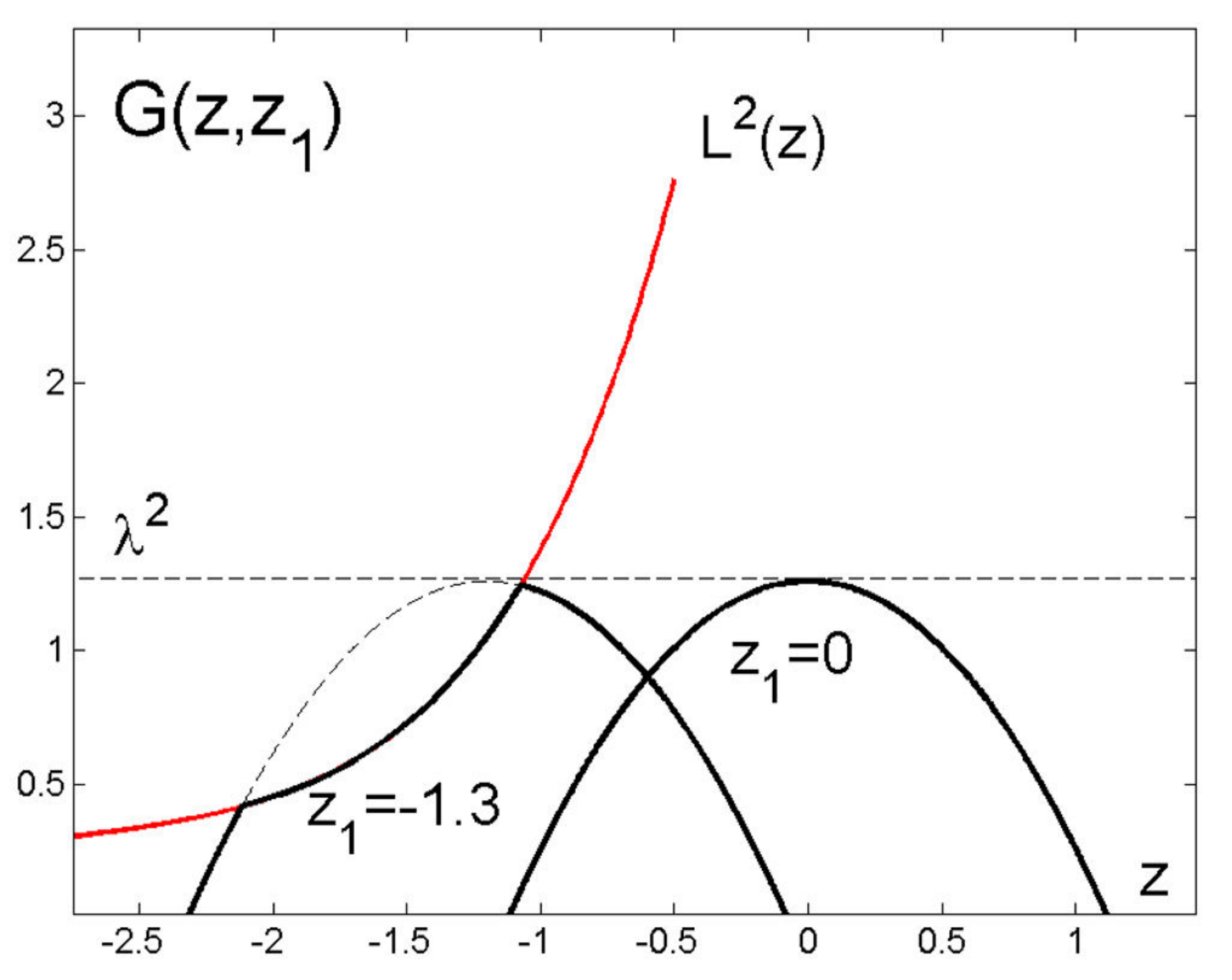}
	\caption{\label{Assym_Kernels} Demonstration of the kernel $G(z,z_1)$ in $F_{att}$ representation (black bold lines) corresponding to two different choice of $z_1$ and square of $\mathcal{L}(z)$ (red line) which devoted to average local geometry properties of random solid surface. Dash line corresponds to behavior of attraction kernels in the absence of surface heterogeneity. }
\end{figure}

\subsection{Hard sphere contribution}

The next term in RSDFT is impact of hard sphere repulsion $F_{HS}$. For this purpose Fundamental Measure Theory (FMT) developed by Rosenfeld \cite{rosenfeld1989free} is used. General expression for Helmholtz free energy of HS has the following form

\begin{eqnarray}
	\label{F_HS_mod}
	F_{HS}=T \int_{0}^{H}S(z)\Phi_{HS}[n_\alpha(z)]dz
\end{eqnarray} 
where as was discussed above spatial integration takes into account not constant area $S(z)$. Several successful modifications of original FMT can be found in literature \cite{roth2010fundamental}. In current research we apply  Rosenfeld-Schmidt-Lowen-Tarazona (RLST) version of  FMT \cite{rosenfeld1997fundamental}. The same version was used in works \cite{neimark2009quenched, ravikovitch2006density} related to adsorption on rough surface. Thus in accordance with RLST $\Phi_{HS}$ can be written as

\begin{eqnarray}
	&\Phi_{HS}[n_\alpha(z)]=-n_0\ln(1-n_3)+\dfrac{n_1n_2-\vec{n}_{v1} \cdot \vec{n}_{v2}}{1-n_3}+ \nonumber \\ 
	&+\dfrac{n_2^3}{24\pi (1-n_3)^2}\left(1-3\left(\dfrac{\vec{n}_{v2}}{n_2}\right)^2+2\left(\dfrac{\vec{n}_{v2}}{n_2}\right)^3\right)
\end{eqnarray}

where $n_{\alpha}$, $\alpha=0,1,2,3$ and $\vec{n}_{\alpha}$, $\alpha=V1, V2$ are scalar and vector averaged functions respectively. These new variables have the following definition:

\begin{eqnarray}
	\label{n_Def}
	&n_{\alpha}(z)=\int d\vec{r'}\rho(z')\omega_{\alpha}(\vec{r}-\vec{r'}), \nonumber
	\\
	& \omega_0(r)=\dfrac{\omega_2(r)}{4\pi R^2},  \,\,\,  \omega_1(r)=\dfrac{\omega_2(r)}{4\pi R},
	\\
	&\omega_2(r)=\delta(R-r), \,\,\, \omega_3(r)=\Theta(R-r),\nonumber
	\\
	&\omega_{V1}(\vec{r})=\dfrac{\omega_{V2}\vec{r}}{4\pi R}, \,\,\, \omega_{V2}\vec{r}=\dfrac{\vec{r}}{r}\delta(R-r) \nonumber
\end{eqnarray}

where $R$ is radius of molecule. As one can see in \eqref{n_Def} the region of integration for $n_{\alpha}$ corresponds to the size of molecule. Situation when a fluid molecule and solid surface has non empty crossover is impossible due to repulsion part of external potential. For this reason expressions \eqref{n_Def} can be used in RSDFT without modifications.

\subsection{External field contribution}

The last term in \eqref{GCE} is external potential describing interaction between a fluid molecule and solid media $F_{ext}$. This potential has to take into account geometrical properties of solid surface. For this reason result of previous section \eqref{external_field} is used. Thus, $F_{ext}$ has the following form:
\begin{eqnarray}
	\label{F_ext}
	F_{ext}=\int dz S(z)\rho(z)U_{fs}(z)
\end{eqnarray}

\subsection{Equlibrium density calculations}

In order to obtain equilibrium density \eqref{Eq_1} one should calculate variations of expressions \eqref{F_id_mod}, \eqref{F_att_mod}, \eqref{F_HS_mod}, \eqref{F_ext}:
\begin{eqnarray}
	\label{var_Fid}
	\dfrac{\delta F_{id}}{\delta \rho}= T S(z) \ln \Lambda \rho(z), \;\;\;\dfrac{\delta F_{ext}}{\delta \rho}= S(z)U_{fs}(z),
\end{eqnarray}
\begin{eqnarray}
	\label{var_Fatt}
	&\dfrac{\delta \beta F_{att}}{\delta \rho}=\dfrac{1}{2}S(z)\int dz'\rho(z')U_{att}G(z, z')+\\ \nonumber
	&+\dfrac{1}{2}\int dz'S(z')\rho(z')U_{att}G(z', z),
\end{eqnarray}
\begin{eqnarray}
	\label{var_Fhs}
	\dfrac{\delta  F_{HS}}{\delta \rho}=T \int dz' S(z') \dfrac{\delta \Phi[n_{\alpha(z')}]}{\delta \rho}
\end{eqnarray}
Important to note that non-symmetry attraction kernel \eqref{Kernel} leads to another deviation in term \eqref{var_Fatt} from the case of smooth surface. Non-symmetry of \eqref{Kernel} is illustrated in Fig.~\ref{Assym_Kernels}.

\begin{table}
	\caption{\label{Table} Parameters of fluid-fluid interaction for argon and nitrogen \cite{neimark2009quenched}, and parameters of solid-fluid interaction for argon and nitrogen on carbon calculated using Lorentz-Berthelot rules \cite{maitland1981}}
	\begin{tabular}{cccccccc}
		\hline
		adsorbate	& $\epsilon_{ff}/k_B$, K&$\sigma_{ff}$, \AA&$\epsilon_{sf}/k_B$, K&  $\sigma_{sf}$, \AA\\
		\hline
		$Ar$ & 111.95 & 3.358 & 56.0 & 3.379 \\
		$N_2$ & 95.77 &  3.549 & 51.8 & 3.475 \\
		\hline			
	\end{tabular}
\end{table}

The adsorption isotherm can be calculated from integration of fluid density distribution. The expression for the number of molecules per unit area has the following form
\begin{eqnarray}
	\label{Ads_Isotherm}
	N_{Ads}=\dfrac{N_A^{-1}}{A}\left(\int_{0}^{z_m}S(z)\rho(z)dz-\rho_0 A z_m\right)
\end{eqnarray}
where $z_m$ is sufficiently large upper limit of integration.
\section{Comparison with Experiment}
\begin{figure*}[tp]
	\includegraphics[width=16.5cm]{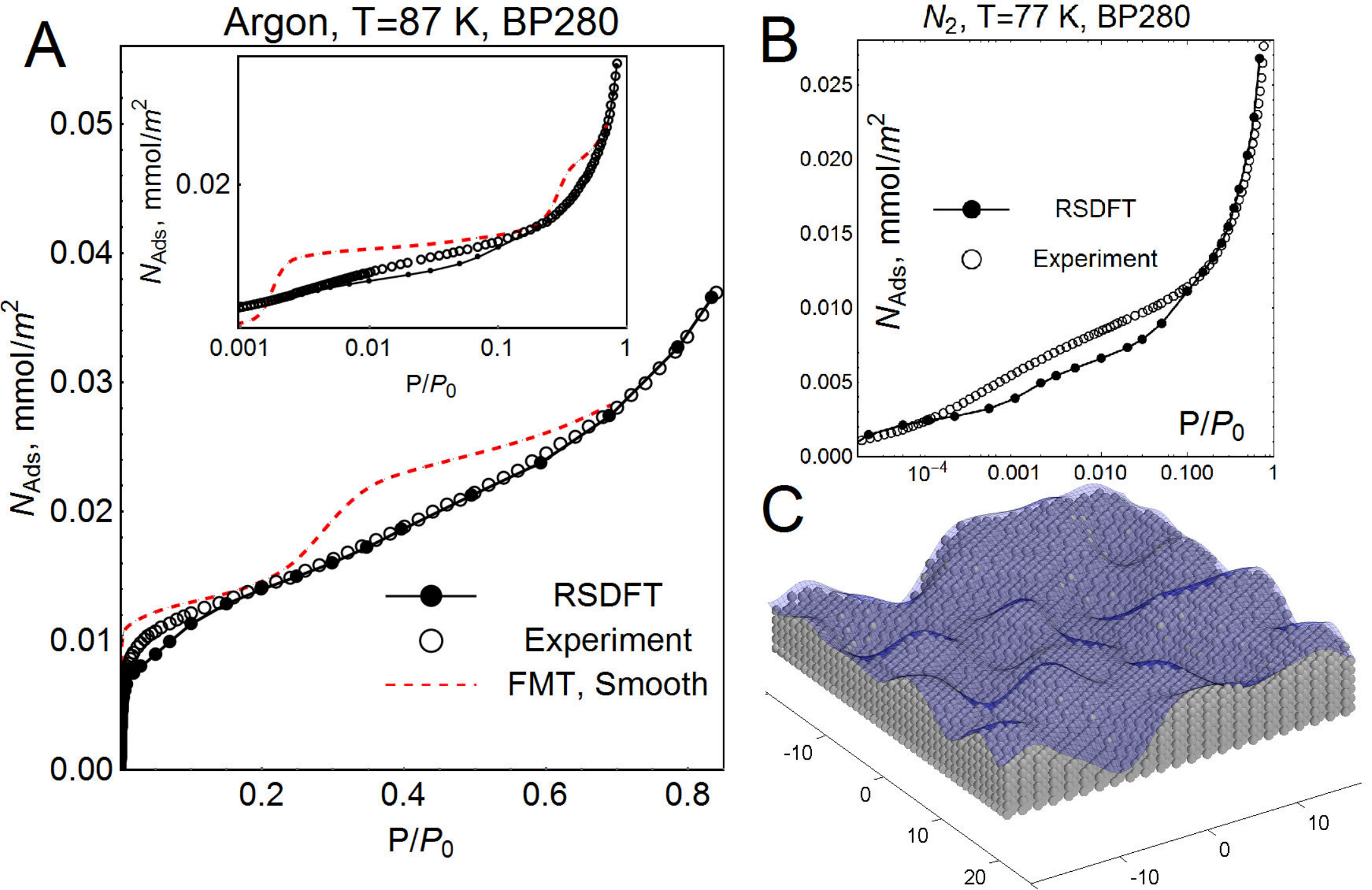}
	\caption{
		(A): \label{ArBP280} Comparison of adsorption isotherms of argon at 87 K on rough surface (BP280 Carbon Black) calculated using RSDFT (black discs joined by solid line) and experimental data \cite{gardner2001reference} (opened circles). RSDFT result corresponds to heterogeneous surface with $\sigma=1.8 d$ and $\tau=5 d$. (B): \label{N2BP280}  Comparison of adsorption isotherms of nitrogen at 77 K on rough surface (BP280 Carbon Black) calculated using RSDFT (black discs joined by solid line) and experimental data \cite{kruk1997nitrogen} (opened circles). RSDFT result corresponds to heterogeneous surface with $\sigma=1.8 d$ and $\tau=5 d$. (C): \label{Surface_1.8_5} Three-dimensional simulated solid media with heterogeneous surface corresponding $\sigma=1.8 d$ and $\tau=5 d$.} 
\end{figure*}
RSDFT developed in current research can be applied for description of adsorption characteristics in case of real heterogeneous surfaces. Ones of the most suitable materials with rough surface are carbon adsorbents of different degree of graphitization. We have considered the argon and nitrogen low temperature adsorption on Cabot BP-280 carbon black. These systems were investigated in a lot of experimental works. This allows us to compare and validate our results obtained by RSDFT with experimental data.

The experiments and the modeling \cite{neimark2009quenched, do2005gcmc} show low temperature adsorption isotherm for smooth surfaces contains steps corresponding to well structured layer formations. Typical adsorption isotherm in the case of ideal surface looks like red dashed line in Fig.~\ref{ArBP280}A.  These types of isotherms can be described well by FMT model with smooth solid wall. However in case of heterogeneous surface there is crucial difference in shape of the isotherms and (as one can see from Fig.~\ref{ArBP280}A) experimental data can not be fitted with standard DFT. For smooth carbon surfaces step-like section of the isotherms starts from around $P/P_0>0.1$. In case of rough surface these steps decrease and then disappear as the surface roughness increases. This effect of isotherm smoothing can not be obtained without taking into spatial heterogeneity of the surface. 
\begin{figure*}[tp]
	\includegraphics[width=16.5cm]{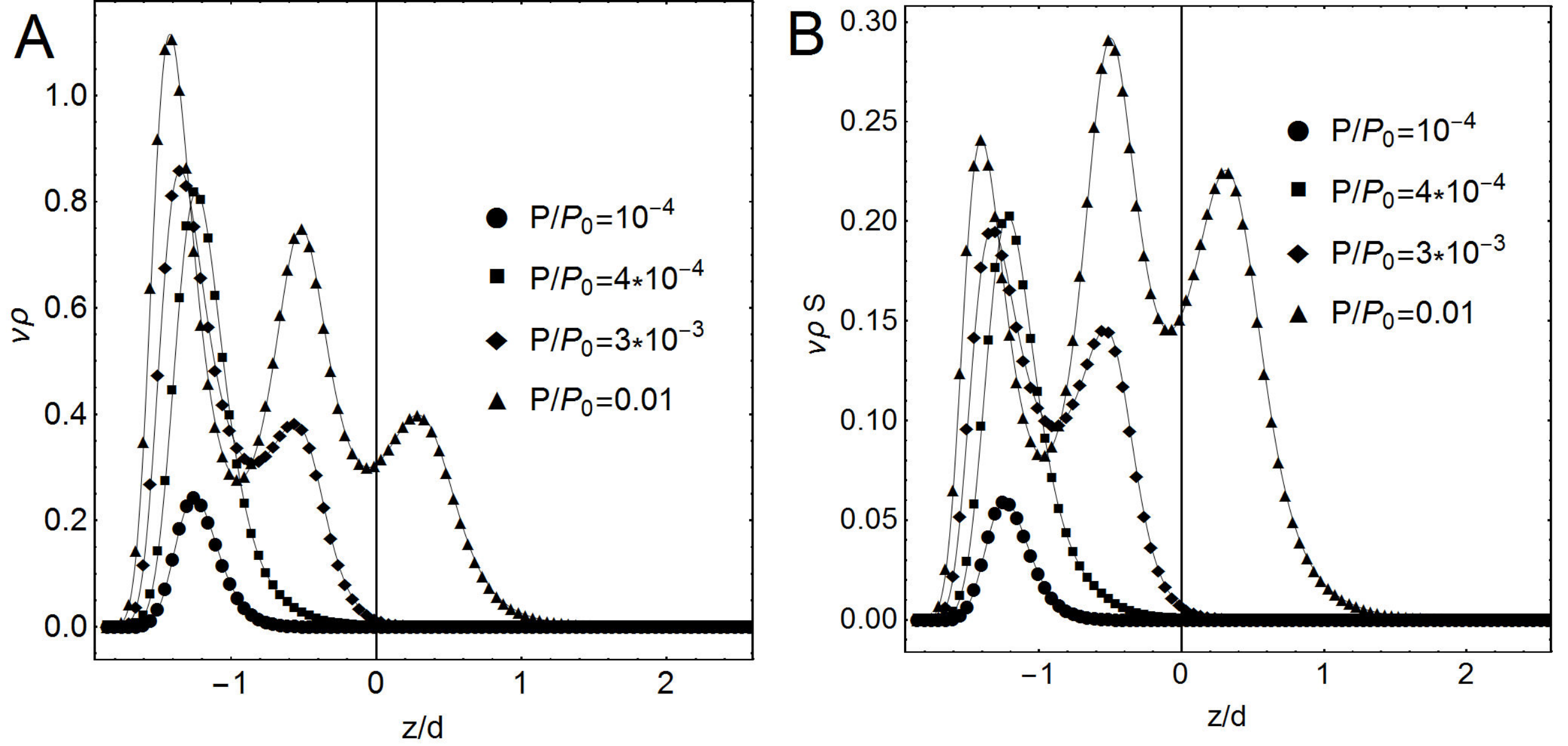}
	\caption{
		(A): \label{DensDistr} Dimensionless density distribution profiles ($\rho v$ where $v$ is molecular volume) corresponding to argon adsorption at 87 K and different pressures on heterogeneous surface with $\sigma=1.8 d$ and $\tau=5 d$. (B): \label{DensDistr_Sz} Density distribution profiles corresponding to argon adsorption at 87 K and different pressures on heterogeneous surface with $\sigma=1.8 d$ and $\tau=5 d$.
	}
\end{figure*}
RSDFT allows to obtain smooth adsorption isotherm for geometrically heterogeneous surfaces. For better demonstration we minimized the number of tunning parameters and fixed characteristics of fluid-solid interaction $\epsilon_{sf}$ and $d_{sf}$. In the case of argon and nitrogen the Lorentz-Berthelot rules are used \cite{maitland1981}: 
\begin{eqnarray}
	\epsilon_{sf}=(\epsilon_{ff}\epsilon_{cc})^{1/2}, \; \; \; \; \;
	d_{sf}=\dfrac{1}{2}(d_{ff}+d_{cc}), \nonumber
\end{eqnarray}
where $\epsilon_{cc}=28/T$ and $d_{cc}=3.4 A$ are characteristic carbon-carbon interaction and carbon molecule diameter respectively\cite{steele1973physical}, corresponding fluid-solid parameters can be found in Table~\ref{Table}. Thus adsorption characteristics are controlled by geometrical properties of the surface: mean square deviation $\sigma$ and correlation length $\tau$. 

For comparison with experimental results we considered different values of roughness parameters $\sigma, \tau$. It was observed that $\sigma=1.8\;d$ and $\tau=5\;d$ provide best agreement with experimental data. Experimental results of low temperature adsorption on BP-280 for argon and nitrogen can be found in \cite{gardner2001reference} and \cite{kruk1997nitrogen}  respectively. 
In Fig.~\ref{ArBP280}A and Fig.~\ref{N2BP280}B circles are the number of adsorbed molecules calculated from experimental data  \cite{gardner2001reference, kruk1997nitrogen} using BET specific surfaces $37 \; m^2 g^{-1}$ and $38 \; m^2 g^{-1}$ for argon and nitrogen respectively \cite{gardner2009argon}. Dashed line corresponds to smooth solid DFT and demonstrates significant deviation from experimental data. 
From Fig.~\ref{ArBP280}A and Fig.~\ref{N2BP280}B one can see that RSDFT results fit well experimental data and show correct smoothed isotherm in range $P/P_0>0.1$.

Correct RSDFT adsorption isotherm for rough surfaces is result of structureless fluid formation near surface. It can be figured out from analysis of fluid density distribution near wall. It is well known that density distribution in case of ideal surface shows distinctly narrow peaks. From the distributions for rough surfaces Fig.\ref{DensDistr}A, one can see that at low pressures argon particles are localized in the defects as these are strong energy sites. As the pressure increases, more layers are formed, but they are not as distinct as those in case of ideal surface due to irregular packing in case of random surface. 

The density $\rho$ defines distribution of molecules inside the space with boundaries defined by $\mathcal{L}(z)$ (Fig.~\ref{Particle_at_z}B). As one can see from expression \eqref{Ads_Isotherm} in order to calculate the number of adsorbed molecules new function  $\bar{\rho}=A^{-1}S(z)\rho(z)$ is needed. Thus, this density controls isotherm smoothness. Indeed one can see from Fig.~\ref{DensDistr_Sz}B with increasing of pressure density distribution profiles change continuously without sharp arising of new peaks. Also boundaries between neighbor peaks are not so obvious and there is competition between the first two peaks with the transition point between $P/P_0=3\times10^{-3}$ and $P/P_0=0.01$. These properties of density distribution lead to lack of step-like region in adsorption isotherm curve.

As one can see RSDFT provides theoretical description of fluid thermodynamic properties near geometric heterogeneous solid. Rigorous approach to stochastic surface description is novel feature of developed model. Also the inverse problem of surface design from experimental investigation can be solved. In Fig.~\ref{Surface_1.8_5}C one can find 3D model of solid wall geometry obtained from experimental isotherms \cite{gardner2001reference, kruk1997nitrogen} matching by RSDFT. For modeling and simulation purposes random surfaces similar tot the one shown in Fig~\ref{Surface_1.8_5} have been generated using the method outlined in \cite{garcia1984monte}, where an uncorrelated distribution of surface points using a random number generator (i.e. white noise) is convolved with  Gaussian filter to achieve correlation.

\section{Non-symmetric models of corrugated materials}
\begin{figure*}[tp] 
	\includegraphics[width=16.5cm]{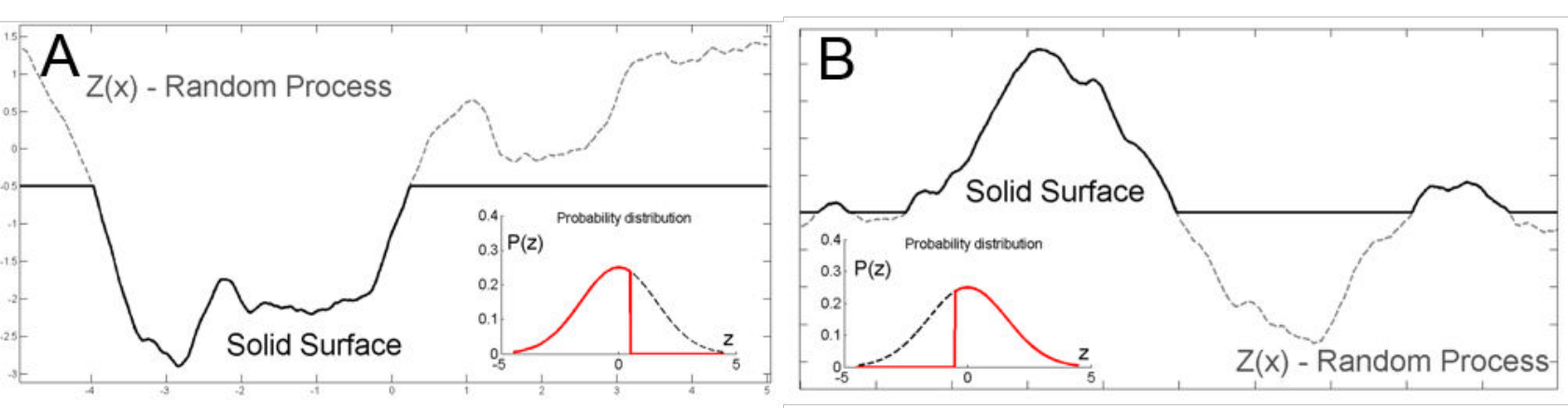}
	\caption{
		(A): \label{Truncation_Up} 
		Illustration of truncation procedure of the random surface from above. Dashed gray line corresponds to original random process realization. Solid black line – final corrugated solid surface in the present model. In the inset probability density distribution is provided in the simplified form (with absence of Dirac delta corresponding to the z-coordinate value of original smooth surface). (B): \label{Truncation_Down} Illustration of truncation procedure of the random surface from below. Dashed gray line corresponds to original random process realization. Solid black line is the final corrugated solid surface in the present model. In the inset probability density distribution is provided in the simplified form (with absence of Dirac delta corresponding to the z-coordinate value of original smooth surface).}
\end{figure*}

\begin{figure*}
	\includegraphics[width=16.5cm]{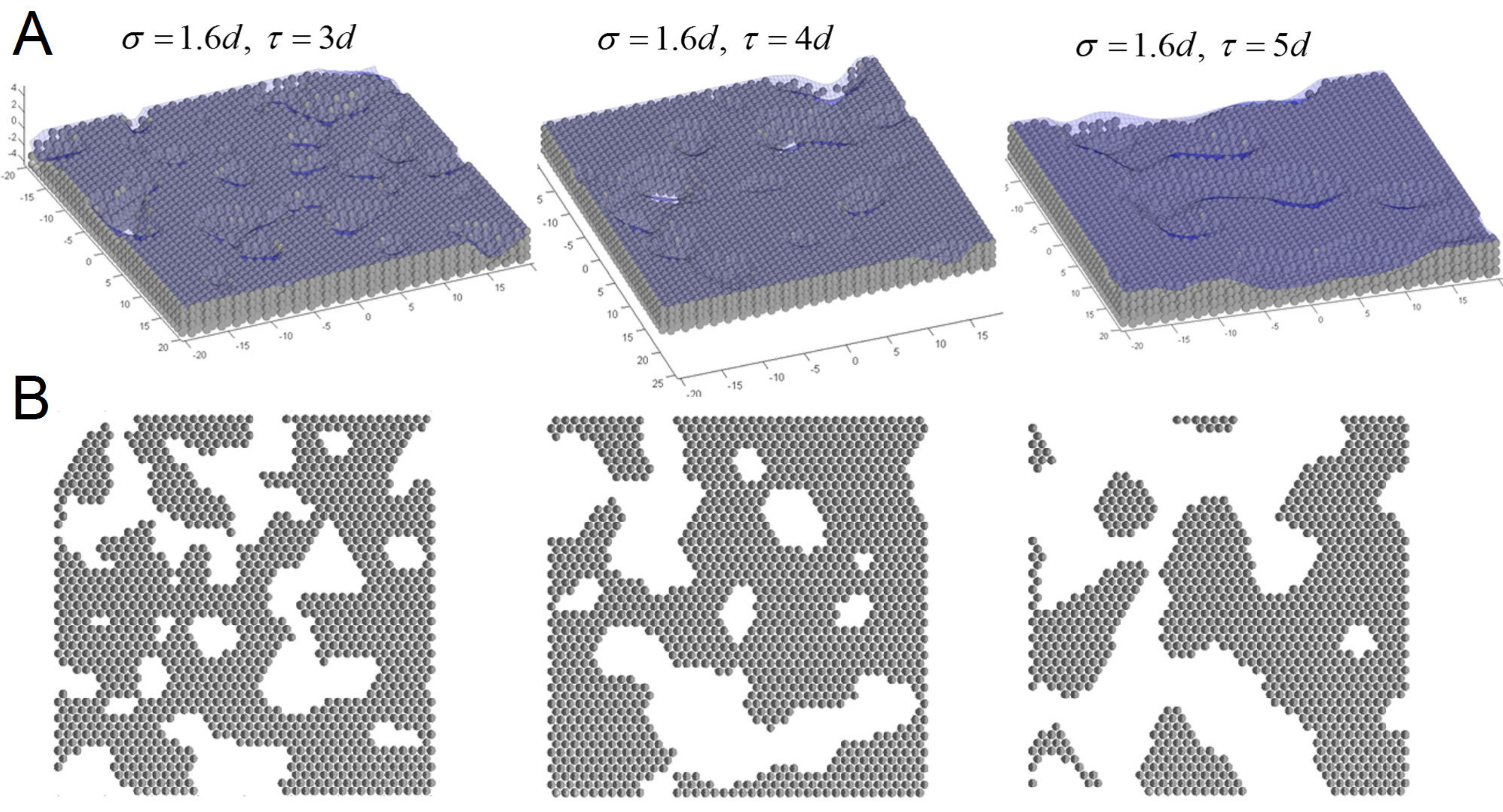}
	\caption{
		(A): \label{Surfaces_Truncation_Up_A}
		3D models of corrugated solid surfaces corresponding to different parameters of random geometry: $\sigma = 1.6$, $\tau =3.0$ (left), $\sigma = 1.6$, $\tau =4.0$ (middle), $\sigma = 1.6$, $\tau =5.0$ (right). (B): \label{Surfaces_Truncation_Up_B} 3D models of corrugated solid surfaces corresponding to different parameters of random geometry:  $\sigma = 1.6$, $\tau =3.0$ (left), $\sigma = 1.6$, $\tau =4.0$ (middle), $\sigma = 1.6$, $\tau =5.0$ (right).}
\end{figure*} 
In this section we begin with discussion of a simple stochastic model for describing a heterogeneous surface which was proposed in  work \cite{do2006modeling}. In spite of its simplicity, this model represents an important stage between the naive approach to the surface stochastic geometry description in the framework of the density functional theory and the complete accurate study of the influence of local properties of random geometry on fluid behavior. The models of heterogeneous surfaces constructed in this manuscript have many similarities to the simple Do-model \cite{do2006modeling} and expand it within the framework of the developed theoretical approach. 

In accordance with work \cite{do2006modeling} heterogeneous surface model contains a collection of graphene layers. Authors of \cite{do2006modeling} considered all layers to be perfect graphene layers, except the top defected one to simplify calculation during fully atomistic GCMC simulation. Also  they noticed  that, in general, defects may be extended to all layers, not only for the top one. To model defects of the top layer, they selected at random a carbon atom in the layer and remove it as well as all surrounding neighbors that have distances to the selected atom less than a certain effective defect radius. This random selection is repeated until the percentage of carbon atoms removed reached a given defect percentage. The percentage of the surface carbon atoms being removed and the effective size of the defect (created by the removal) are the  key parameters to characterize the nongraphitized surface and determine the pattern of isotherm. Once the configuration of the surface has been constructed, the grand canonical Monte Carlo simulation was applied to study the adsorption behavior of argon. As a result authors investigated how various parameters affect the adsorption isotherms. 
\begin{figure*}[tp]
	\includegraphics[width=16.5cm]{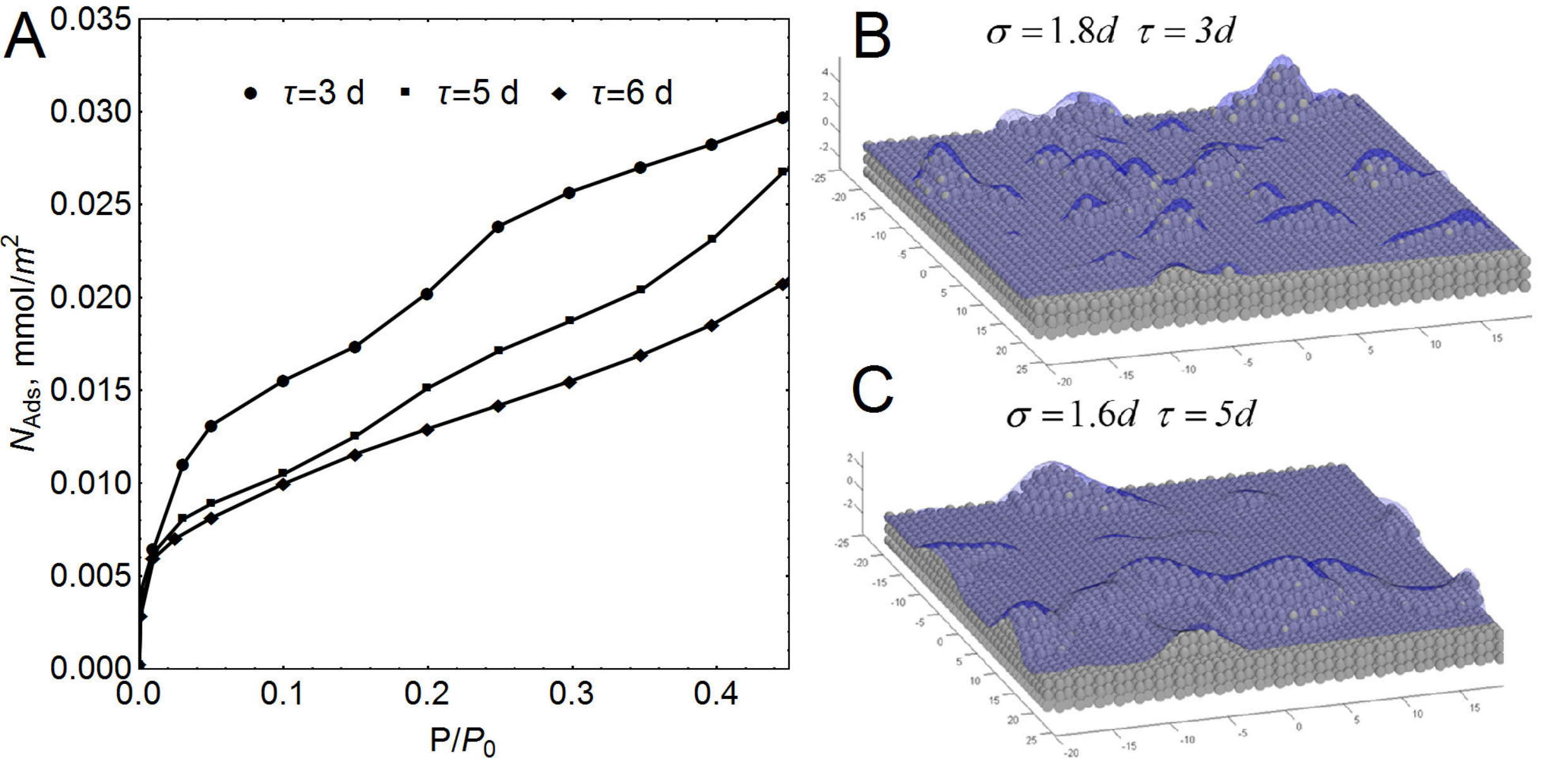}
	\caption{
		(A): \label{CutPotential} Adsorption isotherms calculated by truncated above RSDFT for surfaces with the same $\sigma=1.6$ and different $\tau=3,\;5, \;6$. In all cases fluid is argon at 87 K with parameters from Table~\ref{Table}. (B): \label{Surfaces_truncation_Down}
		3D models obtained by truncation procedure from below. Two models correspond to different parameters of random geometry: $\sigma = 1.8$, $\tau =3.0$, $\sigma =
		1.6$, $\tau =5.0 $.}
\end{figure*}
It is important to note that key parameters which determine the defects of the upper layer and, thus, the adsorption curve, which were introduced in \cite{Khlyupin2017}, completely agree with our approach to describing the geometry and density functional formulation. Let us consider the model of a random process $\mathcal{Z}$ and the characteristics of local geometry described above. Then the percentage of atoms being removed from each z-layer is determined only by the single-point probability distribution function of the process  $\mathcal{Z}$. And the radius of the defects is analogous to the characteristic size of the open subdomains described by function $\mathcal{L}$. Thus, the extended model of corrugated heterogeneous surfaces is obtained in the framework of our approach.

\subsection{Truncation of random process from above }

In the previous section the surface of the solid has been completely described by a random process. It is bounded the solid media from above. 

We have modified the model to describe the process in which random defects are formed in the initially smooth surface. According to this procedure boundary of corrugated solid is a random process immersed inside an initially smooth solid substrate. Also, the regions of solid where the value of random process $\mathcal{Z}$ is bigger than value of initial smooth boundary are cut out. This procedure is called truncation from above.  Sketch illustration of this procedure can be found in Fig.~\ref{Truncation_Up}A

Minor modifications are required in the DFT formulation of the model. One can fully use the previously developed approach taking into account the characteristic jump in the function $\mathcal{L}$, which describes the local geometry (see Fig.~\ref{Particle_at_z}B). Also it is important to note that resulting boundary surface has no longer the Gaussian probability distribution.  Indeed the probability that the height at any point of the surface is  higher than the original height of a smooth substrate equals zero by definition of the truncation procedure (see Fig.~\ref{Truncation_Up}A).

We have applied this truncated DFT to several heterogeneous surface with different $\tau$ parameter. One can see from Fig.~\ref{CutPotential}A our results stay in qualitative agreement with conclusion of\cite{do2006modeling}. Indeed, wavelike behavior of isotherms was observed in the case of perfect flat surface (no defect GTCB). It demonstrates a smoother behavior when the extent of defect is increased and in such case (NGCB) there is no clear plateau of monolayer coverage. This is simply due to the irregular packing in the case of heterogeneous surface geometry, compared to a much more ordered layering in the case of graphitized thermal carbon black. Also, in Fig~\ref{Surfaces_Truncation_Up_A}A and Fig.\ref{Surfaces_Truncation_Up_B}B one can compare relative roughness of surfaces with considered roughness parameters.

\subsection{Truncation of random process from below}

In addition to the surfaces obtained by cutting of random parts from an originally smooth substrate, another model may be considered. Let us imagine that defects with a random form have grown on a original smooth surface. By analogy with the procedure which describe the truncation from above, we introduce the truncation from below to simulate this type of heterogeneous surfaces. All necessary modifications are carried out similarly to the case described above (see Fig.~\ref{Truncation_Down}B for explanations). 3D pictures of these cases can be found in Fig.\ref{Surfaces_truncation_Down}B.

\section{Conclusion}
In this work we have developed new DFT model to describe real fluid behavior near spatial heterogeneous surfaces. The major difference of our result from the existing ones is the stochastic model for solid media, taking into account the correlation properties of the geometry. In order to obtain equilibrium density distribution in the framework of this model we have applied novel effective fluid-solid potential and modified Helmholtz free energy for Lennard-Jones fluid. 

Developed model -- RSDFT is used for calculation of argon and nitrogen low temperature adsorption on real heterogeneous surfaces (BP280 carbon black). The results of RSDFT are in good agreement with experimental data published in the literature. Thus, RSDFT allows one to predict thermodynamical properties of inhomogeneous fluid using essential characteristics of solid wall geometry.

Also the random process theory approach to geometric heterogeneous surface is extended to describe the typical models of corrugated materials. These cases differ in formation nature and correspond to different models of truncated random process. Parametric analysis demonstrates strong influence of surface roughness characteristics on behavior of adsorption isotherms. Therefore competition between impacts of surface heterogeneity and fluid-fluid interactions lead to the variety of isotherm patterns. Presented technique can be used for artificial generation of rough surfaces with desired adsorption characteristics. 

Developed RSDFT may be efficiently applied to several actual problems in surface science: optimal material design for the gas storage problems and double electric layer supercapacitors; influence of spatial heterogeneity on gas-fluid equilibrium, especially in the case of nano-confinement; concurrent adsorption of multicomponent fluid with significant difference in molecular size.

\appendix
\section{Attraction kernels calculation}
In this Appendix we provide explicit expressions of the kernels \eqref{Kernel}. These functions are result of integrations of expression \eqref{F_att} in polar coordinates. As one can see in Fig.~\ref{Particle_at_z} and in the corresponding section local properties of heterogeneous surface are represented by function $\mathcal{L}(z)$. Then integrations in \eqref{F_att} take into account solid media with new boundary defined by  $\mathcal{L}(z)$. It is easy to obtain that there are three possible regions depended on molecule position. These regions correspond to the following conditions: 
\begin{widetext}
The 1st region: $(z_2-z_1)^2\geq\lambda^2$	
\begin{eqnarray}
&G_1(z_1,z_2)=2\pi\int\limits_{0}^{\mathcal{L}(z_2)}rdr\left[\dfrac{C_1}{(\Delta z^2+r^2)^6}+\dfrac{C_2}{(\Delta z^2+r^2)^3}\right] 
\nonumber \\  
&G_1(z_1,z_2)=8 \epsilon_{ff} \pi \left[\dfrac{d^{12}}{10}\left(\dfrac{1}{\Delta z^{10}}-\dfrac{1}{\left(\mathcal{L}(z_2)^{2}+\Delta z^{2}\right)^5}\right)-\dfrac{d^{6}}{4}\left(\dfrac{1}{\Delta z^{4}}-\dfrac{1}{\left(\mathcal{L}(z_2)^{2}+\Delta z^{2}\right)^2}\right)\right]
\end{eqnarray}
The 2nd region: $\lambda^2-\mathcal{L}(z_2)^2<(z_2-z_1)^2<\lambda^2$	
\begin{eqnarray}
&G_2(z_1,z_2)=2\pi \int\limits_0^{\sqrt{\lambda^2-\delta z^2}}(-\epsilon_{ff})rdr+2\pi\int\limits_{\sqrt{\lambda^2-\delta z^2}}^{\mathcal{L}(z_2)}rdr\left[\dfrac{C_1}{(\Delta z^2+r^2)^6}+\dfrac{C_2}{(\Delta z^2+r^2)^3}\right] 
\nonumber \\  
&G_2(z_1,z_2)=-\epsilon_{ff} \pi (\lambda^2+\Delta z^2)+8 \epsilon_{ff} \pi \left[\dfrac{d^{12}}{10}\left(\dfrac{1}{\lambda^{10}}-\dfrac{1}{\left(\mathcal{L}(z_2)^{2}+\Delta z^{2}\right)^5}\right)-\dfrac{d^{6}}{4}\left(\dfrac{1}{\lambda^{4}}-\dfrac{1}{\left(\mathcal{L}(z_2)^{2}+\Delta z^{2}\right)^2}\right)\right]
\end{eqnarray}
The 3rd region: $\mathcal{L}(z_2)^2+(z_2-z_1)^2\leq\lambda^2$
\begin{eqnarray}
G_3(z_1,z_2)=2\pi \int\limits_0^{\mathcal{L}(z_2)}(-\epsilon_{ff})rdr=-\epsilon_{ff} \pi \mathcal{L}(z_2)^2
\end{eqnarray}	

where $\Delta z=z_1-z_2$, $C_1=4 \epsilon_{ff}d^{12}$ and $C_2=4 \epsilon_{ff}d^{6}$
	
\end{widetext}

\bibliography{RandomSurface_DFT}

\begin{thebibliography}{28}%
\makeatletter
\providecommand \@ifxundefined [1]{%
 \@ifx{#1\undefined}
}%
\providecommand \@ifnum [1]{%
 \ifnum #1\expandafter \@firstoftwo
 \else \expandafter \@secondoftwo
 \fi
}%
\providecommand \@ifx [1]{%
 \ifx #1\expandafter \@firstoftwo
 \else \expandafter \@secondoftwo
 \fi
}%
\providecommand \natexlab [1]{#1}%
\providecommand \enquote  [1]{``#1''}%
\providecommand \bibnamefont  [1]{#1}%
\providecommand \bibfnamefont [1]{#1}%
\providecommand \citenamefont [1]{#1}%
\providecommand \href@noop [0]{\@secondoftwo}%
\providecommand \href [0]{\begingroup \@sanitize@url \@href}%
\providecommand \@href[1]{\@@startlink{#1}\@@href}%
\providecommand \@@href[1]{\endgroup#1\@@endlink}%
\providecommand \@sanitize@url [0]{\catcode `\\12\catcode `\$12\catcode
  `\&12\catcode `\#12\catcode `\^12\catcode `\_12\catcode `\%12\relax}%
\providecommand \@@startlink[1]{}%
\providecommand \@@endlink[0]{}%
\providecommand \url  [0]{\begingroup\@sanitize@url \@url }%
\providecommand \@url [1]{\endgroup\@href {#1}{\urlprefix }}%
\providecommand \urlprefix  [0]{URL }%
\providecommand \Eprint [0]{\href }%
\providecommand \doibase [0]{http://dx.doi.org/}%
\providecommand \selectlanguage [0]{\@gobble}%
\providecommand \bibinfo  [0]{\@secondoftwo}%
\providecommand \bibfield  [0]{\@secondoftwo}%
\providecommand \translation [1]{[#1]}%
\providecommand \BibitemOpen [0]{}%
\providecommand \bibitemStop [0]{}%
\providecommand \bibitemNoStop [0]{.\EOS\space}%
\providecommand \EOS [0]{\spacefactor3000\relax}%
\providecommand \BibitemShut  [1]{\csname bibitem#1\endcsname}%
\let\auto@bib@innerbib\@empty
\bibitem [{\citenamefont {Khlyupin}\ and\ \citenamefont
  {Aslyamov}(2017)}]{Khlyupin2017}%
  \BibitemOpen
  \bibfield  {author} {\bibinfo {author} {\bibfnamefont {A.}~\bibnamefont
  {Khlyupin}}\ and\ \bibinfo {author} {\bibfnamefont {T.}~\bibnamefont
  {Aslyamov}},\ }\href {\doibase 10.1007/s10955-017-1786-y} {\bibfield
  {journal} {\bibinfo  {journal} {Journal of Statistical Physics}\ }\textbf
  {\bibinfo {volume} {167}},\ \bibinfo {pages} {1519} (\bibinfo {year}
  {2017})}\BibitemShut {NoStop}%
\bibitem [{\citenamefont {Ravikovitch}\ and\ \citenamefont
  {Neimark}(2006)}]{ravikovitch2006density}%
  \BibitemOpen
  \bibfield  {author} {\bibinfo {author} {\bibfnamefont {P.~I.}\ \bibnamefont
  {Ravikovitch}}\ and\ \bibinfo {author} {\bibfnamefont {A.~V.}\ \bibnamefont
  {Neimark}},\ }\href@noop {} {\bibfield  {journal} {\bibinfo  {journal}
  {Langmuir}\ }\textbf {\bibinfo {volume} {22}},\ \bibinfo {pages} {11171}
  (\bibinfo {year} {2006})}\BibitemShut {NoStop}%
\bibitem [{\citenamefont {Neimark}\ \emph {et~al.}(2009)\citenamefont
  {Neimark}, \citenamefont {Lin}, \citenamefont {Ravikovitch},\ and\
  \citenamefont {Thommes}}]{neimark2009quenched}%
  \BibitemOpen
  \bibfield  {author} {\bibinfo {author} {\bibfnamefont {A.~V.}\ \bibnamefont
  {Neimark}}, \bibinfo {author} {\bibfnamefont {Y.}~\bibnamefont {Lin}},
  \bibinfo {author} {\bibfnamefont {P.~I.}\ \bibnamefont {Ravikovitch}}, \ and\
  \bibinfo {author} {\bibfnamefont {M.}~\bibnamefont {Thommes}},\ }\href@noop
  {} {\bibfield  {journal} {\bibinfo  {journal} {Carbon}\ }\textbf {\bibinfo
  {volume} {47}},\ \bibinfo {pages} {1617} (\bibinfo {year}
  {2009})}\BibitemShut {NoStop}%
\bibitem [{\citenamefont {Jagiello}\ and\ \citenamefont
  {Olivier}(2013)}]{jagiello2013carbon}%
  \BibitemOpen
  \bibfield  {author} {\bibinfo {author} {\bibfnamefont {J.}~\bibnamefont
  {Jagiello}}\ and\ \bibinfo {author} {\bibfnamefont {J.~P.}\ \bibnamefont
  {Olivier}},\ }\href@noop {} {\bibfield  {journal} {\bibinfo  {journal}
  {Adsorption}\ }\textbf {\bibinfo {volume} {19}},\ \bibinfo {pages} {777}
  (\bibinfo {year} {2013})}\BibitemShut {NoStop}%
\bibitem [{\citenamefont {Jagiello}\ \emph {et~al.}(2015)\citenamefont
  {Jagiello}, \citenamefont {Ania}, \citenamefont {Parra},\ and\ \citenamefont
  {Cook}}]{jagiello2015dual}%
  \BibitemOpen
  \bibfield  {author} {\bibinfo {author} {\bibfnamefont {J.}~\bibnamefont
  {Jagiello}}, \bibinfo {author} {\bibfnamefont {C.}~\bibnamefont {Ania}},
  \bibinfo {author} {\bibfnamefont {J.~B.}\ \bibnamefont {Parra}}, \ and\
  \bibinfo {author} {\bibfnamefont {C.}~\bibnamefont {Cook}},\ }\href@noop {}
  {\bibfield  {journal} {\bibinfo  {journal} {Carbon}\ }\textbf {\bibinfo
  {volume} {91}},\ \bibinfo {pages} {330} (\bibinfo {year} {2015})}\BibitemShut
  {NoStop}%
\bibitem [{\citenamefont {Do}\ and\ \citenamefont {Do}(2006)}]{do2006modeling}%
  \BibitemOpen
  \bibfield  {author} {\bibinfo {author} {\bibfnamefont {D.}~\bibnamefont
  {Do}}\ and\ \bibinfo {author} {\bibfnamefont {H.}~\bibnamefont {Do}},\
  }\href@noop {} {\bibfield  {journal} {\bibinfo  {journal} {The Journal of
  Physical Chemistry B}\ }\textbf {\bibinfo {volume} {110}},\ \bibinfo {pages}
  {17531} (\bibinfo {year} {2006})}\BibitemShut {NoStop}%
\bibitem [{\citenamefont {Do}\ and\ \citenamefont {Do}(2005)}]{do2005gcmc}%
  \BibitemOpen
  \bibfield  {author} {\bibinfo {author} {\bibfnamefont {D.}~\bibnamefont
  {Do}}\ and\ \bibinfo {author} {\bibfnamefont {H.}~\bibnamefont {Do}},\
  }\href@noop {} {\bibfield  {journal} {\bibinfo  {journal} {Carbon}\ }\textbf
  {\bibinfo {volume} {43}},\ \bibinfo {pages} {2112} (\bibinfo {year}
  {2005})}\BibitemShut {NoStop}%
\bibitem [{\citenamefont {Qu{\'e}r{\'e}}(2002)}]{quere2002rough}%
  \BibitemOpen
  \bibfield  {author} {\bibinfo {author} {\bibfnamefont {D.}~\bibnamefont
  {Qu{\'e}r{\'e}}},\ }\href@noop {} {\bibfield  {journal} {\bibinfo  {journal}
  {Physica A: Statistical Mechanics and its Applications}\ }\textbf {\bibinfo
  {volume} {313}},\ \bibinfo {pages} {32} (\bibinfo {year} {2002})}\BibitemShut
  {NoStop}%
\bibitem [{\citenamefont {Netz}\ and\ \citenamefont
  {Andelman}(1997)}]{netz1997roughness}%
  \BibitemOpen
  \bibfield  {author} {\bibinfo {author} {\bibfnamefont {R.~R.}\ \bibnamefont
  {Netz}}\ and\ \bibinfo {author} {\bibfnamefont {D.}~\bibnamefont
  {Andelman}},\ }\href@noop {} {\bibfield  {journal} {\bibinfo  {journal}
  {Physical Review E}\ }\textbf {\bibinfo {volume} {55}},\ \bibinfo {pages}
  {687} (\bibinfo {year} {1997})}\BibitemShut {NoStop}%
\bibitem [{\citenamefont {Coasne}\ \emph {et~al.}(2013)\citenamefont {Coasne},
  \citenamefont {Galarneau}, \citenamefont {Pellenq},\ and\ \citenamefont
  {Di~Renzo}}]{coasne2013adsorption}%
  \BibitemOpen
  \bibfield  {author} {\bibinfo {author} {\bibfnamefont {B.}~\bibnamefont
  {Coasne}}, \bibinfo {author} {\bibfnamefont {A.}~\bibnamefont {Galarneau}},
  \bibinfo {author} {\bibfnamefont {R.~J.}\ \bibnamefont {Pellenq}}, \ and\
  \bibinfo {author} {\bibfnamefont {F.}~\bibnamefont {Di~Renzo}},\ }\href@noop
  {} {\bibfield  {journal} {\bibinfo  {journal} {Chemical Society Reviews}\
  }\textbf {\bibinfo {volume} {42}},\ \bibinfo {pages} {4141} (\bibinfo {year}
  {2013})}\BibitemShut {NoStop}%
\bibitem [{\citenamefont {Herminghaus}(2012)}]{herminghaus2012universal}%
  \BibitemOpen
  \bibfield  {author} {\bibinfo {author} {\bibfnamefont {S.}~\bibnamefont
  {Herminghaus}},\ }\href@noop {} {\bibfield  {journal} {\bibinfo  {journal}
  {Physical review letters}\ }\textbf {\bibinfo {volume} {109}},\ \bibinfo
  {pages} {236102} (\bibinfo {year} {2012})}\BibitemShut {NoStop}%
\bibitem [{\citenamefont {Yatsyshin}\ \emph {et~al.}(2017)\citenamefont
  {Yatsyshin}, \citenamefont {Parry}, \citenamefont {Rasc{\'o}n},\ and\
  \citenamefont {Kalliadasis}}]{yatsyshin2017classical}%
  \BibitemOpen
  \bibfield  {author} {\bibinfo {author} {\bibfnamefont {P.}~\bibnamefont
  {Yatsyshin}}, \bibinfo {author} {\bibfnamefont {A.}~\bibnamefont {Parry}},
  \bibinfo {author} {\bibfnamefont {C.}~\bibnamefont {Rasc{\'o}n}}, \ and\
  \bibinfo {author} {\bibfnamefont {S.}~\bibnamefont {Kalliadasis}},\
  }\href@noop {} {\bibfield  {journal} {\bibinfo  {journal} {Journal of
  Physics: Condensed Matter}\ }\textbf {\bibinfo {volume} {29}},\ \bibinfo
  {pages} {094001} (\bibinfo {year} {2017})}\BibitemShut {NoStop}%
\bibitem [{\citenamefont {Persson}\ \emph {et~al.}(2004)\citenamefont
  {Persson}, \citenamefont {Albohr}, \citenamefont {Tartaglino}, \citenamefont
  {Volokitin},\ and\ \citenamefont {Tosatti}}]{persson2004nature}%
  \BibitemOpen
  \bibfield  {author} {\bibinfo {author} {\bibfnamefont {B.}~\bibnamefont
  {Persson}}, \bibinfo {author} {\bibfnamefont {O.}~\bibnamefont {Albohr}},
  \bibinfo {author} {\bibfnamefont {U.}~\bibnamefont {Tartaglino}}, \bibinfo
  {author} {\bibfnamefont {A.}~\bibnamefont {Volokitin}}, \ and\ \bibinfo
  {author} {\bibfnamefont {E.}~\bibnamefont {Tosatti}},\ }\href@noop {}
  {\bibfield  {journal} {\bibinfo  {journal} {Journal of Physics: Condensed
  Matter}\ }\textbf {\bibinfo {volume} {17}},\ \bibinfo {pages} {R1} (\bibinfo
  {year} {2004})}\BibitemShut {NoStop}%
\bibitem [{\citenamefont {Forte}\ \emph {et~al.}(2014)\citenamefont {Forte},
  \citenamefont {Haslam}, \citenamefont {Jackson},\ and\ \citenamefont
  {M{\"u}ller}}]{forte2014effective}%
  \BibitemOpen
  \bibfield  {author} {\bibinfo {author} {\bibfnamefont {E.}~\bibnamefont
  {Forte}}, \bibinfo {author} {\bibfnamefont {A.~J.}\ \bibnamefont {Haslam}},
  \bibinfo {author} {\bibfnamefont {G.}~\bibnamefont {Jackson}}, \ and\
  \bibinfo {author} {\bibfnamefont {E.~A.}\ \bibnamefont {M{\"u}ller}},\
  }\href@noop {} {\bibfield  {journal} {\bibinfo  {journal} {Physical Chemistry
  Chemical Physics}\ }\textbf {\bibinfo {volume} {16}},\ \bibinfo {pages}
  {19165} (\bibinfo {year} {2014})}\BibitemShut {NoStop}%
\bibitem [{\citenamefont {Ustinov}\ \emph {et~al.}(2006)\citenamefont
  {Ustinov}, \citenamefont {Do},\ and\ \citenamefont
  {Fenelonov}}]{ustinov2006pore}%
  \BibitemOpen
  \bibfield  {author} {\bibinfo {author} {\bibfnamefont {E.}~\bibnamefont
  {Ustinov}}, \bibinfo {author} {\bibfnamefont {D.}~\bibnamefont {Do}}, \ and\
  \bibinfo {author} {\bibfnamefont {V.}~\bibnamefont {Fenelonov}},\ }\href@noop
  {} {\bibfield  {journal} {\bibinfo  {journal} {Carbon}\ }\textbf {\bibinfo
  {volume} {44}},\ \bibinfo {pages} {653} (\bibinfo {year} {2006})}\BibitemShut
  {NoStop}%
\bibitem [{\citenamefont {Ustinov}\ \emph {et~al.}(2005)\citenamefont
  {Ustinov}, \citenamefont {Do},\ and\ \citenamefont
  {Jaroniec}}]{ustinov2005application}%
  \BibitemOpen
  \bibfield  {author} {\bibinfo {author} {\bibfnamefont {E.}~\bibnamefont
  {Ustinov}}, \bibinfo {author} {\bibfnamefont {D.}~\bibnamefont {Do}}, \ and\
  \bibinfo {author} {\bibfnamefont {M.}~\bibnamefont {Jaroniec}},\ }\href@noop
  {} {\bibfield  {journal} {\bibinfo  {journal} {Applied surface science}\
  }\textbf {\bibinfo {volume} {252}},\ \bibinfo {pages} {548} (\bibinfo {year}
  {2005})}\BibitemShut {NoStop}%
\bibitem [{\citenamefont {Khlyupin}(2016)}]{khlyupin2016effects}%
  \BibitemOpen
  \bibfield  {author} {\bibinfo {author} {\bibfnamefont {A.}~\bibnamefont
  {Khlyupin}},\ }in\ \href@noop {} {\emph {\bibinfo {booktitle} {Journal of
  Physics: Conference Series}}},\ Vol.\ \bibinfo {volume} {774}\ (\bibinfo
  {organization} {IOP Publishing},\ \bibinfo {year} {2016})\ p.\ \bibinfo
  {pages} {012024}\BibitemShut {NoStop}%
\bibitem [{\citenamefont {Wu}(2006)}]{wu2006density}%
  \BibitemOpen
  \bibfield  {author} {\bibinfo {author} {\bibfnamefont {J.}~\bibnamefont
  {Wu}},\ }\href@noop {} {\bibfield  {journal} {\bibinfo  {journal} {AIChE
  Journal}\ }\textbf {\bibinfo {volume} {52}},\ \bibinfo {pages} {1169}
  (\bibinfo {year} {2006})}\BibitemShut {NoStop}%
\bibitem [{\citenamefont {Kalikmanov}(2013)}]{kalikmanov2013statistical}%
  \BibitemOpen
  \bibfield  {author} {\bibinfo {author} {\bibfnamefont {V.}~\bibnamefont
  {Kalikmanov}},\ }\href@noop {} {\emph {\bibinfo {title} {Statistical physics
  of fluids: basic concepts and applications}}}\ (\bibinfo  {publisher}
  {Springer Science \& Business Media},\ \bibinfo {year} {2013})\BibitemShut
  {NoStop}%
\bibitem [{\citenamefont {Rosenfeld}(1989)}]{rosenfeld1989free}%
  \BibitemOpen
  \bibfield  {author} {\bibinfo {author} {\bibfnamefont {Y.}~\bibnamefont
  {Rosenfeld}},\ }\href@noop {} {\bibfield  {journal} {\bibinfo  {journal}
  {Physical review letters}\ }\textbf {\bibinfo {volume} {63}},\ \bibinfo
  {pages} {980} (\bibinfo {year} {1989})}\BibitemShut {NoStop}%
\bibitem [{\citenamefont {Roth}(2010)}]{roth2010fundamental}%
  \BibitemOpen
  \bibfield  {author} {\bibinfo {author} {\bibfnamefont {R.}~\bibnamefont
  {Roth}},\ }\href@noop {} {\bibfield  {journal} {\bibinfo  {journal} {Journal
  of Physics: Condensed Matter}\ }\textbf {\bibinfo {volume} {22}},\ \bibinfo
  {pages} {063102} (\bibinfo {year} {2010})}\BibitemShut {NoStop}%
\bibitem [{\citenamefont {Rosenfeld}\ \emph {et~al.}(1997)\citenamefont
  {Rosenfeld}, \citenamefont {Schmidt}, \citenamefont {L{\"o}wen},\ and\
  \citenamefont {Tarazona}}]{rosenfeld1997fundamental}%
  \BibitemOpen
  \bibfield  {author} {\bibinfo {author} {\bibfnamefont {Y.}~\bibnamefont
  {Rosenfeld}}, \bibinfo {author} {\bibfnamefont {M.}~\bibnamefont {Schmidt}},
  \bibinfo {author} {\bibfnamefont {H.}~\bibnamefont {L{\"o}wen}}, \ and\
  \bibinfo {author} {\bibfnamefont {P.}~\bibnamefont {Tarazona}},\ }\href@noop
  {} {\bibfield  {journal} {\bibinfo  {journal} {Physical Review E}\ }\textbf
  {\bibinfo {volume} {55}},\ \bibinfo {pages} {4245} (\bibinfo {year}
  {1997})}\BibitemShut {NoStop}%
\bibitem [{\citenamefont {Maitland}\ \emph {et~al.}(1981)\citenamefont
  {Maitland}, \citenamefont {Rigby}, \citenamefont {Smith},\ and\ \citenamefont
  {Wakeham}}]{maitland1981}%
  \BibitemOpen
  \bibfield  {author} {\bibinfo {author} {\bibfnamefont {G.}~\bibnamefont
  {Maitland}}, \bibinfo {author} {\bibfnamefont {M.}~\bibnamefont {Rigby}},
  \bibinfo {author} {\bibfnamefont {E.}~\bibnamefont {Smith}}, \ and\ \bibinfo
  {author} {\bibfnamefont {W.}~\bibnamefont {Wakeham}},\ }\href@noop {}
  {\enquote {\bibinfo {title} {Intermolecular forces},}\ } (\bibinfo {year}
  {1981})\BibitemShut {NoStop}%
\bibitem [{\citenamefont {Gardner}\ \emph {et~al.}(2001)\citenamefont
  {Gardner}, \citenamefont {Kruk},\ and\ \citenamefont
  {Jaroniec}}]{gardner2001reference}%
  \BibitemOpen
  \bibfield  {author} {\bibinfo {author} {\bibfnamefont {L.}~\bibnamefont
  {Gardner}}, \bibinfo {author} {\bibfnamefont {M.}~\bibnamefont {Kruk}}, \
  and\ \bibinfo {author} {\bibfnamefont {M.}~\bibnamefont {Jaroniec}},\
  }\href@noop {} {\bibfield  {journal} {\bibinfo  {journal} {The Journal of
  Physical Chemistry B}\ }\textbf {\bibinfo {volume} {105}},\ \bibinfo {pages}
  {12516} (\bibinfo {year} {2001})}\BibitemShut {NoStop}%
\bibitem [{\citenamefont {Kruk}\ \emph {et~al.}(1997)\citenamefont {Kruk},
  \citenamefont {Jaroniec},\ and\ \citenamefont {Gadkaree}}]{kruk1997nitrogen}%
  \BibitemOpen
  \bibfield  {author} {\bibinfo {author} {\bibfnamefont {M.}~\bibnamefont
  {Kruk}}, \bibinfo {author} {\bibfnamefont {M.}~\bibnamefont {Jaroniec}}, \
  and\ \bibinfo {author} {\bibfnamefont {K.~P.}\ \bibnamefont {Gadkaree}},\
  }\href@noop {} {\bibfield  {journal} {\bibinfo  {journal} {Journal of Colloid
  and Interface Science}\ }\textbf {\bibinfo {volume} {192}},\ \bibinfo {pages}
  {250} (\bibinfo {year} {1997})}\BibitemShut {NoStop}%
\bibitem [{\citenamefont {Steele}(1973)}]{steele1973physical}%
  \BibitemOpen
  \bibfield  {author} {\bibinfo {author} {\bibfnamefont {W.~A.}\ \bibnamefont
  {Steele}},\ }\href@noop {} {\bibfield  {journal} {\bibinfo  {journal}
  {Surface Science}\ }\textbf {\bibinfo {volume} {36}},\ \bibinfo {pages} {317}
  (\bibinfo {year} {1973})}\BibitemShut {NoStop}%
\bibitem [{\citenamefont {Gardner}\ and\ \citenamefont
  {Jaroniec}(2009)}]{gardner2009argon}%
  \BibitemOpen
  \bibfield  {author} {\bibinfo {author} {\bibfnamefont {L.}~\bibnamefont
  {Gardner}}\ and\ \bibinfo {author} {\bibfnamefont {M.}~\bibnamefont
  {Jaroniec}},\ }\href@noop {} {\bibfield  {journal} {\bibinfo  {journal}
  {Annales UMCS, Chemistry}\ }\textbf {\bibinfo {volume} {64}},\ \bibinfo
  {pages} {49} (\bibinfo {year} {2009})}\BibitemShut {NoStop}%
\bibitem [{\citenamefont {Garcia}\ and\ \citenamefont
  {Stoll}(1984)}]{garcia1984monte}%
  \BibitemOpen
  \bibfield  {author} {\bibinfo {author} {\bibfnamefont {N.}~\bibnamefont
  {Garcia}}\ and\ \bibinfo {author} {\bibfnamefont {E.}~\bibnamefont {Stoll}},\
  }\href@noop {} {\bibfield  {journal} {\bibinfo  {journal} {Physical review
  letters}\ }\textbf {\bibinfo {volume} {52}},\ \bibinfo {pages} {1798}
  (\bibinfo {year} {1984})}\BibitemShut {NoStop}%
\end{thebibliography}%
\end{document}